\documentclass[sn-mathphys,Numbered]{sn-jnl}

\usepackage{graphicx}%
\usepackage{multirow}%
\usepackage{amsmath,amssymb,amsfonts}%
\usepackage{amsthm}%
\usepackage{mathrsfs}%
\usepackage[title]{appendix}%
\usepackage{xcolor}%
\usepackage{textcomp}%
\usepackage{manyfoot}%
\usepackage{booktabs}%
\usepackage{algorithm}%
\usepackage{algorithmicx}%
\usepackage{algpseudocode}%
\usepackage{listings}%
\usepackage{subfig}

\theoremstyle{thmstyleone}%
\theoremstyle{thmstyletwo}%

\theoremstyle{thmstylethree}%

\begin{document}

\title[Article Title]{Effect of the Surface Dimples on the Exit Dynamics of a Sphere at a Constant Velocity}


\author*[1]{\fnm{Intesaaf} \sur{Ashraf}} \email{intesaaf.ashraf15@gmail.com}

\author[1]{\fnm{Stephane} \sur{ Dorbolo}} \email{s.dorbolo@uliege.be}
\equalcont{These authors contributed equally to this work.}

\affil*[1]{\orgdiv{GRASP, Institute of Physics}, \orgname{University of Liege}, \orgaddress{\city{Liege}, \postcode{4000}, \country{Belgium}}}


\abstract{This article experimentally investigates the exit dynamics of two different spheres i.e. a smooth sphere and a sphere with dimples at a constant speed.  Employing high-speed experimental observations, the study investigates the key characteristic lengths and hydrodynamic forces between the sphere's surface and water.  The experiments were performed at different traveling speeds (or Froude numbers).  The results shed light on the drag force coefficient, entrainment force coefficient, and cross-over force coefficient. In this paper, we have shown that the surface dimples not only affect the drag force but also the entrainment force. However, the cross-over force coefficient remains the same.}


\maketitle

\section{Introduction}\label{sec1}

The exit dynamics of an object is an interesting fluid mechanics problem. The early research on exit dynamics was carried out by Havelock (\cite{Havelock1936}, \cite{havelock1949resistance}, \cite{havelock1949wave}), in which the impulsively starting motion of a sphere was studied using the linear theory, with constant velocity and constant acceleration, respectively. 
Greenhow and Lin \cite{greenhow1983nonlinear} carried out fundamental studies of non-linear free surface effects both experimentally and theoretically at constant velocity.  Further, Greenhow and Lin \cite{greenhow1983nonlinear} have shown that the upward movement caused an elevation of the free surface that looks like a bump. This elevation or bump chaotically breaks down, which is termed waterfall breaking. 

Telste \cite{telste1987} computationally studied the exit of a sphere using potential flow theory. In this study, it is shown that at low speed ($Fr^2<0.04$), the sphere approach resembles as it is moving against a wall. Whereas at the high speed ($Fr^2>2$), while approaching the free surface, the sphere resembles as if it was moving in an infinite fluid.  

In the literature, we can also find the work of Nair et. al. \cite{nair2018water}  in which an improved VOF model was implemented to study the exit dynamics of a sphere from water. 
Chu et. al. \cite{chu2010} studied the water exit of a sphere. In this work, they observed cavities at both ends of the sphere. The slapping of the water was reported because of the collapse of those cavities. They performed the experiments for both the accelerating and decelerating motion of the sphere. Buruchenko \& Canelas \cite{buruchenko2017validation} proposed a new model based on Smoothed Particle Hydrodynamics and studied the entry and exit of a two-dimensional cylinder. 

Haohao et. al. \cite{haohao2019numerical} carried out the  LBM (Lattice Boltzmann methods) simulations of the exit dynamics of a sphere at a constant velocity. They have shown, a strong dependency of free surface elevation on Froude number below 4.12. The waterfall breaking is a function of the Froude number. On the other hand, the Reynolds number is the dominant parameter when the sphere moves beneath the water's surface. Ni et. al. \cite{ni2015} through the simulation of a fully submerged spheroid water exit and demonstrated that the moment at which the free surface breaks up from the body can be delayed by making the object blunter. 

Truscott et. al \cite{truscott2016} experimentally studied the behavior of a buoyant sphere that pops out to the water surface under the action of the buoyancy. They have shown that the vortical signature of the sphere is highly dependent upon the Reynolds number. 
The trajectory of the sphere can be either a straight line or oscillating depending upon the Reynolds number.
Wu et al. \cite{wu2017experimental} experimentally observed that for a fully submerged sphere, the height of water elevation increases with the Froude number. In the case of a partially submerged sphere, the water drains out from the sphere surface at the end of the exit stage as a water column stage. The height of the water column increases with the increase in Froude number.

After going through all these literature surveys, we can notice that all the studies were carried out for the smooth sphere. There is no information about how surface roughness or texture affects the exit dynamics of the sphere. In this study, we have entered a new frontier with the exploration of dimpled spheres. Dimples, those small depressions on a surface, have long been recognized for their aerodynamic benefits in the context of golf balls \cite{mehta1985aerodynamics, choi2006mechanism}. The transition from water to air, however, brings a fresh perspective to the role of dimples in exit dynamics. In this study, two non-dimensionalized numbers are varied in this experimental study i.e. Froude number and the Reynolds number. 

\section{Experimental Setup}
\begin{figure}
	\centering
	\includegraphics[scale=0.50]{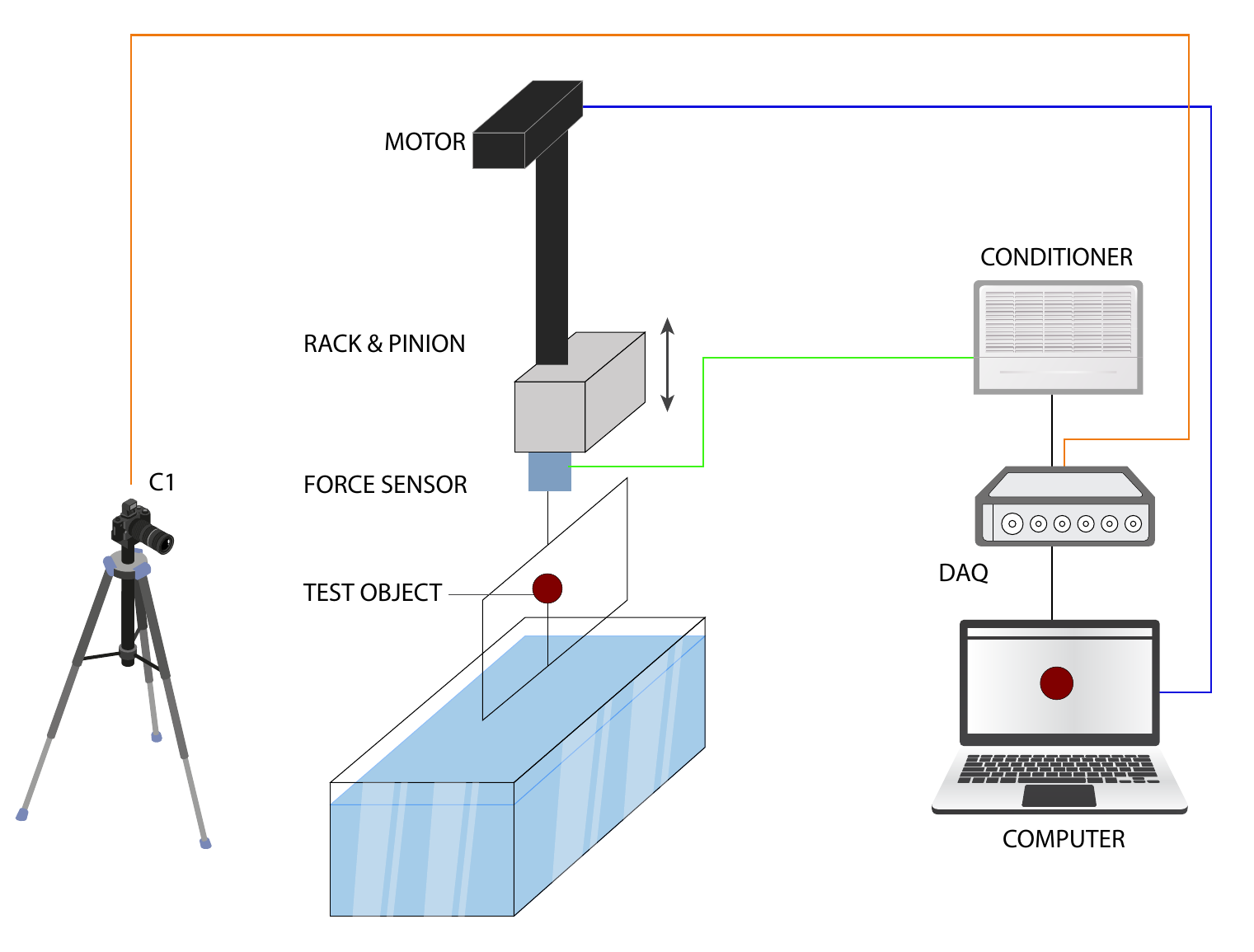}
	\caption{The schematic of Experimental Set-up. The sphere was screwed into a carbon fiber frame. The frame was attached to the force whereas the force sensor was attached to the rack and pinion mechanism. C1 was a high-speed camera used to record the motion.}
	\label{setup}
\end{figure}

The experimental setup is shown in Fig. \ref{setup}. It consists of rack and pinion mechanics which were attached to a frame, water tank, sphere, and a high-speed camera. The sphere was attached to the frame. The force sensor was positioned between the frame and the rack and pinion mechanism.  The experimental setup was previously used and described by Ashraf et. al. \cite{ashraf2023exit} in their experimental study of the exit dynamics of a square a sphere. The experiments were carried out for two different spheres i.e. smooth sphere (diameter, $a=$ 65 mm and density, $\rho=$139 kg/m$^3$) and a sphere with dimples like a golf ball (diameter, $a=$ 71 mm and density, $\rho=$110 kg/m$^3$). The sphere with dimples has  4 mm dia circular dimples, which have a depth of 0.5 mm and a pitch of 7 mm. The dimples are distributed uniformly all over the ball.  Both balls are commercial hockey balls. 

The sphere started moving from rest and at a depth of $y=-$30 cm ($y=0$ was set when the center of mass of the sphere is located at the water-air interface at rest) from the water surface. The sphere motion started with a constant acceleration of 4 m/s$^2$. Therefore, a certain traveling distance was needed to reach the set-point constant velocity, e.g. 12.5 cm is required to reach a vertical velocity of 1 m/s. 


\section{Result and Discussion}
\begin{figure}
	\centering
	\subfloat[][]{\includegraphics[width=0.75\textwidth]{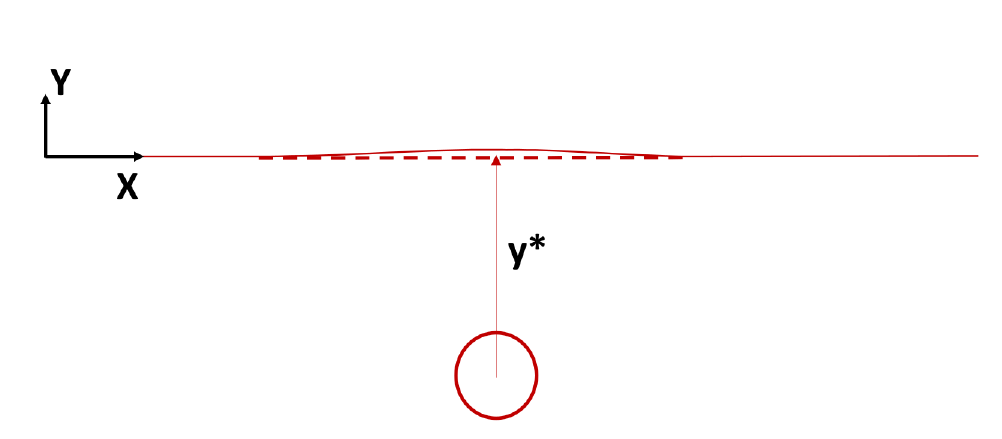}}\\
	\subfloat[][]{\includegraphics[width=0.75\textwidth]{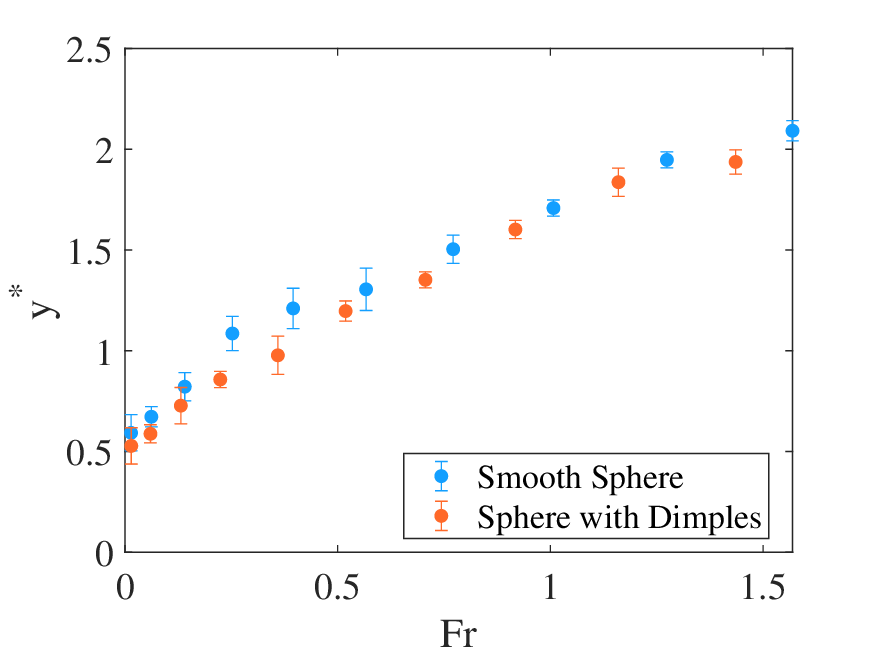}}
	\caption{(a) Scheme illustrating the maximum  depth of deformation, $y^*$.  (b) $y^*$ as a function of the Froude number. The symbols are the mean value of 7 experiments and the error bars are standard deviation.}
	\label{fig:1}       
\end{figure}

As the sphere moves upward, it deforms the free surface. The surface presents a bump whose size increases when the sphere gets closer and closer to the top. We can define the depth $y_1$ for which the bump elevation is $10\%$ of sphere radius.  Figure \ref{fig:1} shows $y^*=y_1/a$ as a function of the Froude number for the two spheres with and without dimples.  In both cases, $y^*$ increases with an increase in the Froude Number for both spheres which is expected. Moreover, the data for both spheres exhibit the same trends. We can conclude that the observation of the surface deformation cannot discriminate if the surface is textured or not.

\begin{figure}
	\centering
	\subfloat[][]{\includegraphics[width=0.75\textwidth]{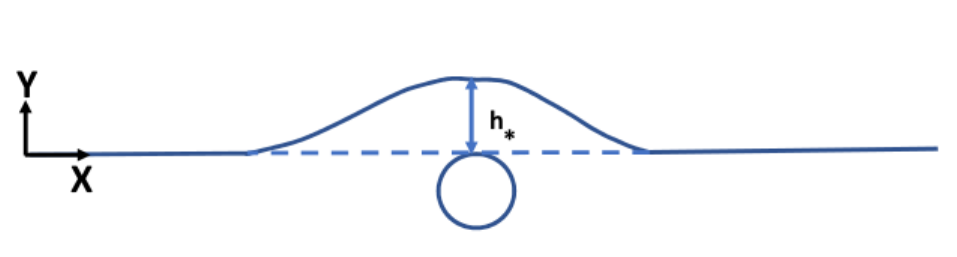}}\\
	\subfloat[][]{\includegraphics[width=0.75\textwidth]{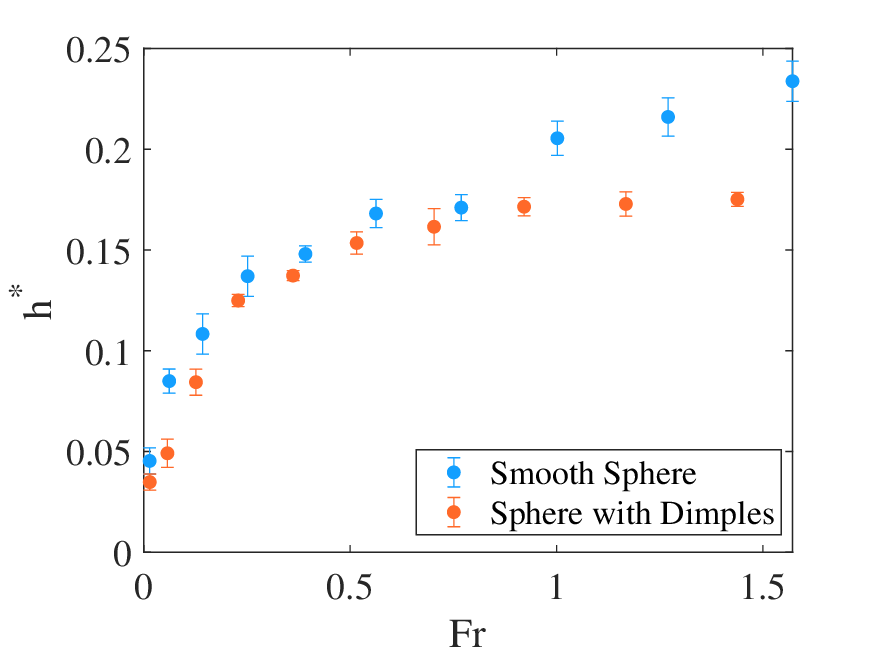}}

	\caption{ (a) Scheme illustrating the maximum deformation of the free surface, $h^*$. (b) $h^*$ as a function of Froude number. The symbols are the mean value of 7 experiments and the error bars are standard deviation.}
	\label{fig:2}       
\end{figure}

The bump deformation can be measured by image analysis. We define $h_0$ as the deformation size of the bump when the sphere is located at $y=-a/2$. Fig. \ref{fig:2} shows the maximum elevation, $h^*=h_0/a$, achieved by the sphere as a function of the Froude number. Similarly to $y^*$, the value of $h^*$ first increases with an increase in the Froude number without any surprise. However, above $Fr=0.8$, the results obtained for the smooth sphere and the dimpled sphere are separated. The $h^*$,  in the case of a smooth sphere,  is higher above $Fr=0.8$. 

To understand this change of behavior at high Froude, we measured the force needed to keep the speed constant as a function of the position of the sphere. The forces acting on the sphere during its upward movement and interaction with the free water surface are studied using a stain gauge sensor. 

\begin{figure}
	\centering
	\includegraphics[width=0.75\textwidth]{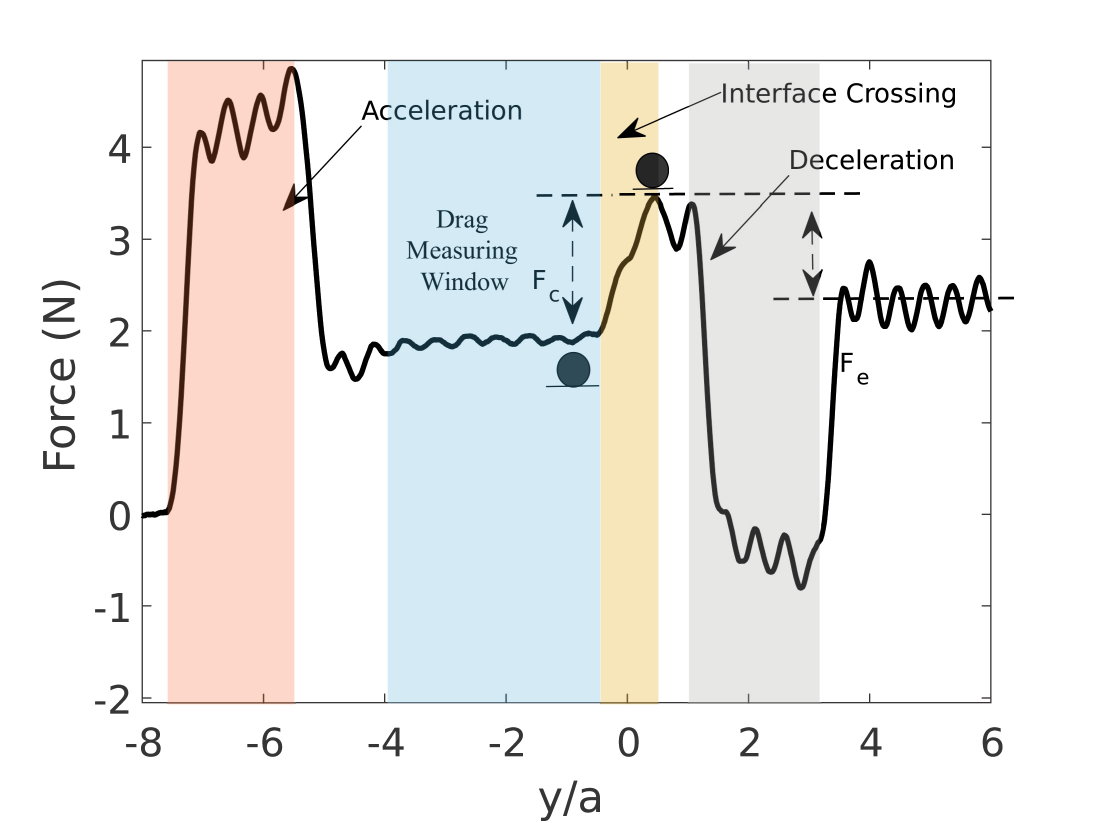}\\
	\caption{ Sample force as a function of $y/a$ position for sphere with dimples at $Fr=0.71$. It illustrates the different force regimes acting on the sphere during its upward motion.}
	\label{fig:4}       
\end{figure}

\begin{figure}
	\centering
	\includegraphics[width=0.75\textwidth]{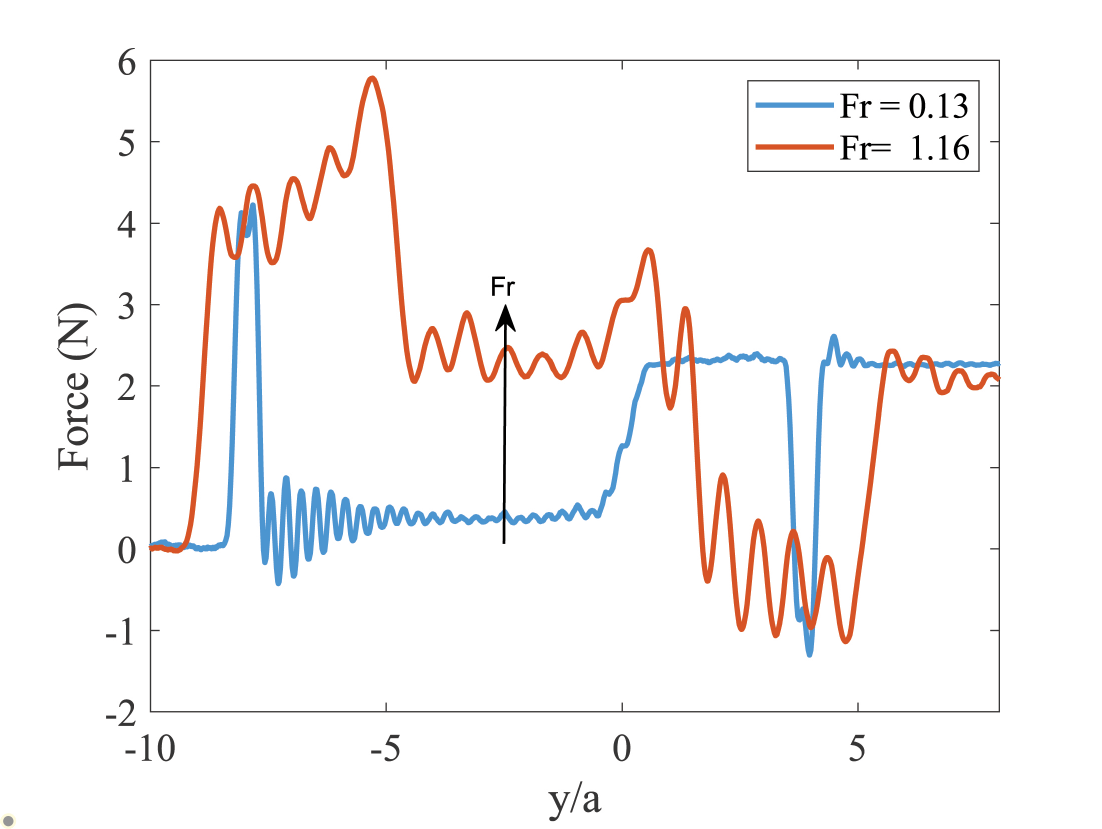}\\
	\caption{ Force as a function of $y/a$ position for a sphere with dimples at different Froude numbers.}
	\label{fig:7}       
\end{figure}

\begin{figure}
	\centering
	\subfloat[][]{\includegraphics[width=.75\textwidth]{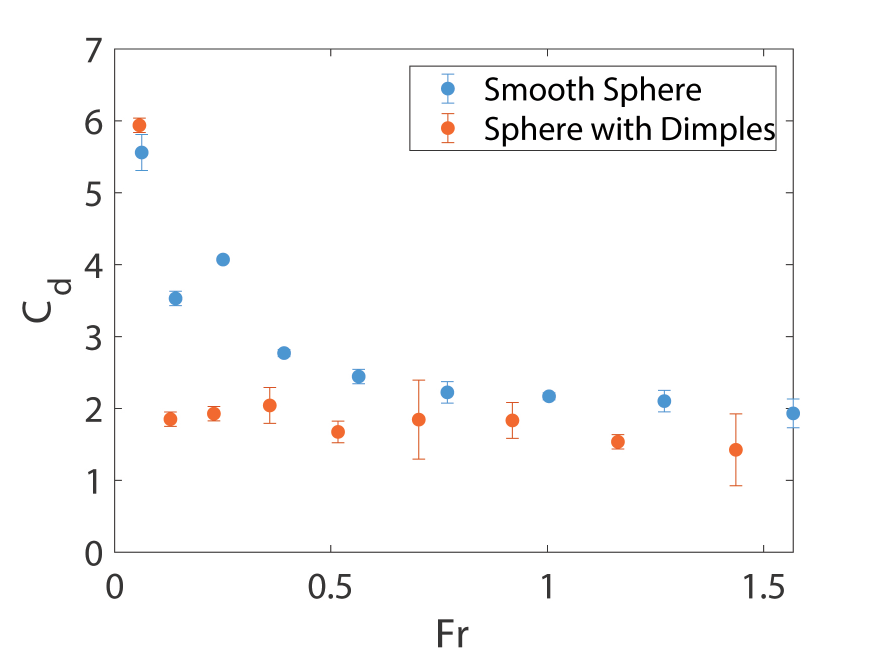}}\\
	\subfloat[][]{\includegraphics[width=.75\textwidth]{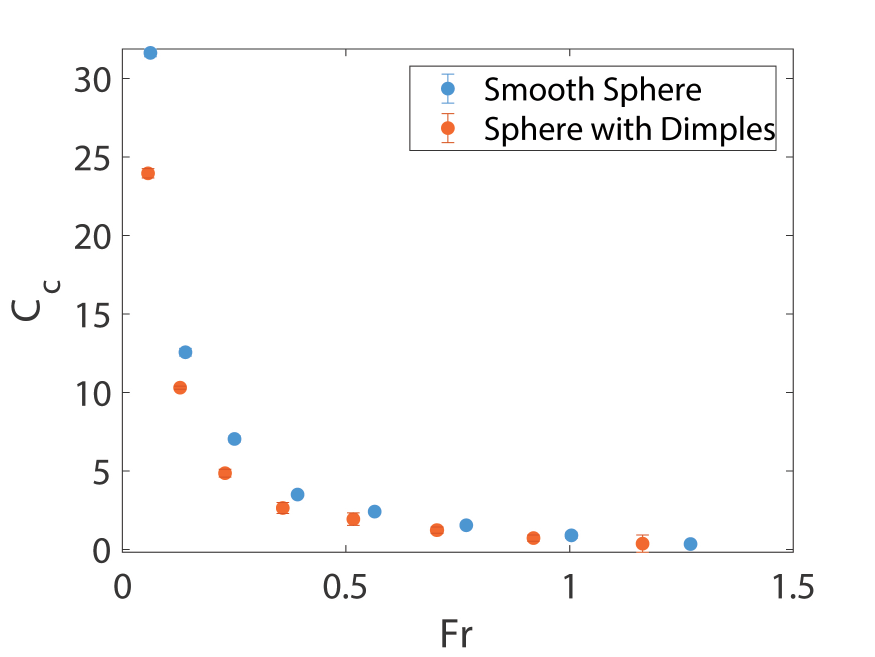}}\\
	\subfloat[][]{\includegraphics[width=.75\textwidth]{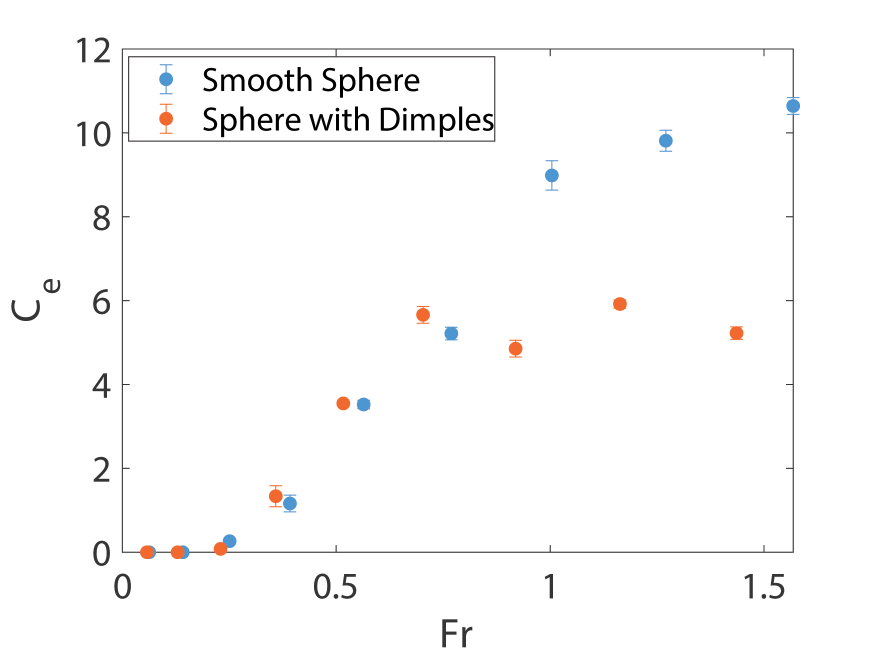}}
	\caption{ Force Coefficients as a function of Froude number. (a)Drag Coefficient, (b) Crossing-over force Coefficient, and (c) Entrainment force Coefficient. The symbols are the mean value of 3 experiments and error bars are standard deviations.   }
	\label{fig:3}       
\end{figure}

Fig. \ref{fig:4} shows the registered force variation as a function of depth for a sphere with dimples at Fr = 0.71. The sphere motion is categorized into 5 zones or stages, as shown in the graph: 
(a) The acceleration phase: It is highlighted in orange. The sphere moves from rest to acquire the constant velocity priory set for each experiment. At the end of this stage, the sphere starts moving upward at a constant speed.
(b) The drag force regime.  It is highlighted in blue.  The sphere moves vertically in water at a constant velocity. The sphere experiences the drag force in this stage. Just at the end of the acceleration stage, the drag force is quite high, and then it settles down to relatively constant values as can be seen by the flat plateau.  During this flat plateau regime, the drag force $F_d$  was measured. The net drag force is measured by taking the average force acting on the sphere during this period.  
(c) The crossing phase, highlighted in yellow. It is also a constant velocity regime.  The sphere starts crossing the interface when its top is in first contact with the free water surface and the crossing ends when the bottom of the sphere leaves the water surface.  The crossing-over force $F_c$ is measured by taking the difference of force at those spheres' positions i.e. $y/a=-0.5$ and $y/a=0.5$.  
(d) Finally, the sphere comes out of the water. The entrainment force $F_e$ is measured when $y/a=0.5$ minus the weight out of the water. It is equal to the force acting on the sphere when it is completely out of the water minus the weight of the sphere when it stops moving after the deceleration stage. The deceleration stage is depicted in gray color.  
These regimes are similar to what has been reported in the literature for the exit dynamics of a square cylinder (see Ashraf et. al \cite{ashraf2023exit}).
The force regime for a sphere with dimples as a function of Froude numbers is shown in Fig. \ref{fig:7}. It shows that with the increase in Froude number, the drag force and entrainment force increase whereas the crossing-over force decreases. The crossing-over force is a balance between the buoyancy, components of surface tension in the vertical directions, drag force, and weight of the sphere. While other forces remain constant as a function of the Froude number, the drag force however increases with the increase in Froude number. It results in a decrease in the net force acting on the sphere when it crosses the interface.   Fig. \ref{fig:3} shows the mean drag coefficient, interface crossing over force coefficient, and entrainment coefficient acting on the sphere.  The drag force and crossing-over force coefficients have been calculated by dividing the net force by $0.5 \rho A U^2$, where $A$ is the cross-sectional area of the sphere, $\rho$ is the density of water and U is the vertical velocity of the sphere. 

Fig. \ref{fig:3}a show that the drag force coefficient is lower for spheres with dimples for the smooth sphere, especially at a high Froude number. It is not a new phenomenon but a well-known fact that the surface dimples reduce the drag force acting on the sphere \cite{beratlis2019origin}. This reduction in drag is due to the dimples on the surface which causes the boundary layer to transition from laminar to turbulent flow \cite{mehta1985aerodynamics, choi2006mechanism}. By creating this turbulent boundary layer, the separation point decreases. The dimpled ball experiences less separation and thus less drag. 


From Fig. \ref{fig:3}b,  It can be seen that the crossing-over coefficient decreases with the increase in the Froude number. It is the coefficient of the net force acting on the sphere while crossing the interface.   The crossing-over force coefficients are similar for both spheres. Therefore, the surface dimples do not affect crossing over the interface. 

Another interesting value is the entrainment coefficient, as shown in Fig. \ref{fig:3}c. It is calculated by using the following steps: 
\begin{itemize}
	\item Calculate the Entrained Force ($F_e$): We measured it from the force plot as shown in Fig. \ref{fig:4}.
	\item Calculate the Volume of the Sphere ($V_{sphere}$):
	The volume of the sphere is a characteristic property of the object. For a sphere, the volume can be calculated using the formula: $V_{sphere} = \frac{4}{3} \pi r^3$, Where $r$ is the radius of the sphere.
	\item Relate the Entrained Force to the Volume of Water displaced by the sphere during entrainment:  
	The volume of water displaced by the sphere when the cylinder is out of the water is calculated by the formula, $V_{displaced} =  \frac{F_e}{\rho_{water}}$. Where, $V_{displaced}$ is the volume of water displaced by the sphere, $F_e$ is the entrained force acting on the sphere, and $\rho_{water}$ is the density of the water. It's important to note that this transformation assumes that the entrained force is primarily due to buoyancy and that the volume of displaced water is directly related to the volume of the submerged part of the sphere. Additionally, other forces (e.g. surface tension) effects are considered negligible in the analysis.
	\item Calculate the Coefficient of entrainment ($C_e$):
	It is calculated  by using the formula   $C_e = \frac{V_{displaced}}{V_{sphere}} $.
\end{itemize}

It can be that at a low Froude number, there is very little entrainment. However, as the Froude increases ($Fr > 0.21$), the entrainment starts. Now with an increase in the Froude number, the entrainment coefficient increases. Another thing to be noted is that the entrainment coefficient is similar for a sphere with dimples as compared to a smooth sphere up to the Froude Number, $Fr \approx 0.8$. However, after the Froude number $Fr \approx 0.8$, the entrainment coefficient is higher for a smooth sphere, which is similar to $h^*$ results. It means that there is less entrained volume in the case of a sphere with dimples. The reason for less entrainment again can be the transition from a laminar boundary layer to a turbulent boundary layer in the case of dimples. The dimples convert the boundary layer into a turbulent one. It means the wake size decreases in case of turbulent boundary layer \cite{lignarolo2011shape}. As a result, less fluid will be dragged along by the wake of the sphere and so less fluid will be entrained by the sphere. 

\section{Conclusion}
Finally, from all the results above, we can conclude that as soon as the sphere moves upward at a constant speed, it starts elevating the free surface water. The depth at which water surface elevation starts is a function of the Froude number. Also, the maximum surface elevation is achieved, when the top of the sphere touches the water surface. However, this maximum surface elevation increases with the Froude number and represents the entrainment of the water. The sphere with dimples has considerably lower entrainment as compared to the smooth sphere at a higher Froude number. As a result, it has a lower entrainment force coefficient too. The drag coefficient is also lower for spheres with dimples, which is a well-known phenomenon. This means surface dimples help the sphere in exiting the fluid surface. A sphere with dimples popping out of the water surface or fluid will pop out to a higher distance because of the lower drag force and entrainment force coefficient as compared to the smooth sphere.  The crossing-over coefficient is found to be similar for both spheres. 
\section*{Declarations}
\subsection*{Ethical Approval} This declaration is “not applicable”.
\subsection*{Funding} 
This work was financially supported by PDR WOLFLOW (F.R.S. -- FNRS) and by Fonds Sp\'eciaux de la Recherche (ULiege). SD is an FNRS senior research associate.
\subsection*{Availability of data and materials} The data can be available on reasonable request from the corresponding author. 


\end{document}